\documentstyle[psfig,aps,multicol,eqsecnum]{revtex}
\begin{document}

\title{Asymmetric quantum cloning machines in any dimension}

\author{Nicolas J. Cerf$^{1,2,3}$}
\address{$^1$W. K. Kellogg Radiation Laboratory,
California Institute of Technology, Pasadena, California 91125\\
$^2$Information and Computing Technologies Research Section,
Jet Propulsion Laboratory, Pasadena, California 91109\\
$^3$Center for Nonlinear Phenomena and Complex Systems,
Universit\'e Libre de Bruxelles, 1050 Bruxelles, Belgium}

\date{May 1998, revised August 1998}

\draft
\maketitle

\begin{abstract}

A family of asymmetric cloning machines for $N$-dimensional quantum
states is introduced. These machines produce two imperfect 
copies of a single state that emerge from two distinct
Heisenberg channels. The tradeoff between the quality of these copies
is shown to result from a complementarity akin to Heisenberg
uncertainty principle. A no-cloning inequality is
derived for isotropic cloners:
if $\pi_a$ and $\pi_b$ are the depolarizing fractions associated 
with the two copies, the domain in $(\sqrt{\pi_a},\sqrt{\pi_b})$-space
located inside a particular ellipse representing close-to-perfect
cloning is forbidden. More generally,
a no-cloning uncertainty relation is discussed,
quantifying the impossibility of copying imposed by quantum mechanics.
Finally, an asymmetric
Pauli cloning machine is defined that makes two approximate copies
of a quantum bit,  while the input-to-output operation underlying
each copy is a (distinct) Pauli channel. 
The class of symmetric Pauli cloning machines is shown
to provide an upper bound on the quantum capacity
of the Pauli channel of probabilities $p_x$, $p_y$ and $p_z$.
The capacity is proven to be vanishing if
$(\sqrt{p_x},\sqrt{p_y},\sqrt{p_z})$ lies outside an ellipsoid
whose pole coincides with the universal cloning machine.

\end{abstract}
\pacs{PACS numbers: 03.67.Hk, 03.65.Bz, 89.70.+c
      \hfill KRL preprint MAP-224}


\section{Introduction}

A fundamental property of quantum information is that
it cannot be copied, in contrast with information we are used to
in classical physics.
This means that there exists no physical process that can produce
perfect copies of a system that is initially in an 
{\em unknown} quantum state. This so-called {\em no-cloning} theorem,
recognized by Dieks~\cite{bib_dieks} and Wootters and Zurek~\cite{bib_wz},
is an immediate consequence of the linearity of quantum mechanics,
and lies at the heart of quantum theory. Remarkably,
if cloning {\em was} permitted, the Heisenberg uncertainty principle
could then be violated by measuring conjugate observables on many copies
of a single quantum system. Nevertheless, even if {\em perfect} 
quantum cloning is precluded by the uncertainty principle, 
it is possible to devise a cloning machine that yields
{\em imperfect} copies of the original quantum state, that is,
the cloning process necessarily introduces errors.
Quantum cloning machines have recently attracted a lot of attention
because of their use in connection with quantum communication and
cryptography (see, e.g.,~\cite{bib_gisinhuttner,bib_bruss}).
\par

To fix the ideas, consider a cloning machine that duplicates 
a two-state system (a quantum bit or a qubit) that is initially
in an arbitrary state $|\psi\rangle=\alpha |0\rangle + \beta |1\rangle$
whose amplitudes $\alpha$ and $\beta$ are unknown. 
It is easy to build a cloning machine that perfectly copies 
the two basis states $|0\rangle$ and $|1\rangle$, 
but then it badly duplicates the superpositions 
$2^{-1/2}(|0\rangle\pm |1\rangle)$.
In other words, it cannot produce perfect copies of
{\em all} possible input states simultaneously.
This being so, we may ask how well
one can {\em approximately} duplicate the unknown state 
of the qubit if the quality of the copies is required
to be independent of the input state. This question has been answered 
by Buzek and Hillery~\cite{bib_bh} who first showed that
it is possible to construct a cloning machine
that yields two imperfect copies of a single qubit in state $|\psi\rangle$.
Specifically, they defined a universal cloning machine (UCM)
which creates two identical copies of $|\psi\rangle$
characterized each by the same density
operator $\rho=(2/3)|\psi\rangle\langle\psi|+\openone/6$.
This machine is called {\em universal} because it produces copies that 
are {\em state-independent}: the fidelity of cloning
$f\equiv\langle\psi|\rho|\psi\rangle=5/6$ does not depend on
the input state. Equivalently, the two output qubits of the UCM can be
viewed as emerging from a depolarizing channel of probability 1/4, 
that is, a channel whose errors are {\em isotropic} (the Bloch vector
characterizing the state $|\psi\rangle$ is shrunk by a factor
2/3 regardless its orientation). The UCM was later proved
to be optimal by Bruss et al.~\cite{bib_bruss}
and Gisin and Massar~\cite{bib_gisinmassar}
when the quality of a cloner is measured by the fidelity $f$.
The concept of approximate cloning was also extended
by Gisin and Massar who considered optimal $m$-to-$n$ cloners,
which produce $m$ imperfect copies from $n$ identical original qubits.
More recently, the optimal cloning of quantum bits was generalized
to quantum systems of arbitrary dimensions by Buzek and 
Hillery~\cite{bib_bh_registers} ($1$-to-$2$ universal cloner), and
Werner~\cite{bib_werner} ($m$-to-$n$ optimal cloners).
Zanardi~\cite{bib_zanardi} also discussed a group-theoretical
analysis of $m$-to-$n$ optimal cloners for pure states
of arbitrary dimension.
\par

In this paper, we introduce a family of {\em asymmetric} cloning machines
that produce two non-identical (approximate) copies of the state of an
$N$-dimensional quantum system. This is in contrast with the cloning
machines considered so far, which are symmetric (the copies being identical,
or characterized by the same density operator).\footnote{A recent study
of asymmetric and anisotropic cloning of qubits has been carried out
independently by Niu and Griffiths~\cite{bib_niu}. It was
pointed out to us after completion of this work.}
The two copies created by our cloners emerge from quantum channels 
that are defined with a basis of $N^2$ error operators 
on $N$-dimensional inputs that form a Heisenberg group 
(see~\cite{bib_fivel}). We refer to these channels
as {\em Heisenberg} channels, and the corresponding cloners
as {\em Heisenberg} cloning machines (HCM).
A Heisenberg channel is characterized by the $N^2$-dimensional
probability distribution of the error operators 
that the quantum state undergoes in the channel.
Using a particular class of {\em isotropic} asymmetric HCMs,
i.e., cloners whose outputs emerge from (distinct) depolarizing channels,
we derive a {\em no-cloning inequality} governing 
the tradeoff between the quality of the copies of a single 
$N$-dimensional state imposed by quantum mechanics:
\begin{equation}  \label{eq_1.1}
a^2+2ab/N+b^2\ge 1 \;, 
\end{equation}
where $\pi_a=a^2$ and $\pi_b=b^2$ are the depolarizing fractions of the
channels associated with outputs $A$ and $B$, respectively.
It is a tight inequality for all isotropic cloning machines,
which is saturated with the HCMs.
This ellipse in ($a,b$)-space tends to a circle when copying
$N$-dimensional states with $N\to\infty$, 
which has a simple semi-classical interpretation.
More generally, the complementarity between the two copies
produced by an anisotropic HCM
is shown to result from a genuine {\em uncertainty principle}, 
much like that associated with Fourier transforms.
Accordingly, the probability distributions ${\mathbf p}$ and ${\mathbf q}$
characterizing
the Heisenberg channels leading to the two outputs of the cloner
cannot be peaked simultaneously, giving rise to a balance
between the quality of the copies. This can be expressed by an
entropic no-cloning uncertainty relation, 
$H({\mathbf p})+H({\mathbf q}) \ge \log_2(N^2)$, where $H(\cdot)$
stands for the Shannon entropy.
\par

In Section~\ref{sect_qubits}, we discuss
the asymmetric cloning of two-dimensional quantum states,
in order to prepare the grounds of the extension to $N$ dimensions.
We introduce a {\em Pauli} cloning machine (PCM)~\cite{bib_cerf}, which
produces two (generally non-identical) output qubits,
each emerging from a {\em Pauli} channel. A Pauli channel 
is a special case ($N=2$) of a Heisenberg channel which is
defined by the four-element group of error operators for qubits,
generated by the bit/phase flip errors (see Sec.~\ref{sect_paulichannel}).
The family of PCMs relies on a parametrization of 4-qubit wave functions
for which all qubit pairs are in a mixture of Bell states.
The resulting no-cloning uncertainty relation for qubits is discussed.
In particular, the subclass of {\em isotropic} asymmetric PCMs 
is used in order to derive a tight no-cloning inequality for qubits, i.e.,
Eq.~(\ref{eq_1.1}) for $N=2$.
The subclass of {\em symmetric} PCMs is then used to
express an upper bound on the quantum capacity of a Pauli channel,
generalizing the considerations of Bruss et al.~\cite{bib_bruss}
for a depolarizing channel. In particular, the capacity 
of the Pauli channel with probabilities $p_x=x^2$, $p_y=y^2$ and $p_z=z^2$,
is shown to be vanishing if $(x,y,z)$ lies outside the ellipsoid
\begin{equation}
x^2+y^2+z^2+xy+xz+yz=1/2
\end{equation}
whose pole coincides with the depolarizing channel
that underlies the UCM. Assuming that $C$ is a continuous function
of $p_x$, $p_y$, and $p_z$, this yields an upper bound on the capacity,
namely $C \le 1-2(x^2+y^2+z^2+xy+xz+yz)$.
\par

In Section~\ref{sect_Ndim}, we generalize the Pauli cloning machine 
to systems of arbitrary dimensions, and define a family of asymmetric 
Heisenberg cloning machines for $N$-dimensional states. 
Our description is based on
the $N^2$ maximally-entangled states of two $N$-dimensional systems
which generalize the Bell states, and the corresponding Heisenberg
group of error operators in $N$ dimensions~\cite{bib_fivel}.
The family of asymmetric HCMs is used
to investigate the complementarity principle governing
the tradeoff between the quality of the copies.
The optimal $N$-dimensional UCM~\cite{bib_bh_registers,bib_werner}
is shown to be a special case (symmetric and isotropic)
of these cloners. 
\par

\section{Pauli cloning machines for quantum bits}
\label{sect_qubits}

\subsection{Characterization of a Pauli channel using Bell states}
\label{sect_paulichannel}

Consider a quantum bit in an arbitrary state 
$|\psi\rangle=\alpha |0\rangle + \beta |1\rangle$
which is processed by a Pauli channel~\cite{bib_cerf}. 
A Pauli channel is defined using the group of four
error operators (the three Pauli matrices $\sigma_{x,y,z}$ and 
the identity $\openone$), namely it acts on state $|\psi\rangle$
by either rotating it by one of the Pauli matrices or leaving it unchanged.
Specifically, the input qubit undergoes 
a phase-flip ($\sigma_z$), a bit-flip ($\sigma_x$), 
or their combination ($\sigma_x \sigma_z=-i\sigma_y$) with
respective probabilities $p_z$, $p_x$, and $p_y$, or
remains unchanged with probability $p=1-p_x-p_y-p_z$.
A depolarizing channel corresponds to the special case where $p_x=p_y=p_z$.
It is very convenient to describe the operation of a Pauli channel
by considering an input qubit $X$ maximally entangled 
with a reference qubit $R$. Indeed, a remarkable property of entanglement
is that by applying a unitary transformation on just one qubit 
of a two-qubit system, one can transform a maximally-entangled
joint state into another.
Denoting the four maximally-entangled states of two qubits (or Bell states) as
\begin{equation}
|\Phi^{\pm}\rangle = {1\over\sqrt{2}} (|00\rangle \pm |11\rangle)
\qquad  \qquad
|\Psi^{\pm}\rangle = {1\over\sqrt{2}} (|01\rangle \pm |10\rangle)
\end{equation}
it is easy to check that the {\em local} action of the error operators
on one of them, say $|\Phi^+\rangle$, yields the three
remaining Bell states, namely\footnote{Note that
we use the convention $|0\rangle=|\uparrow\rangle$
and $|1\rangle=|\downarrow\rangle$.}
\begin{eqnarray}  \label{eq_actionPauli}
(\openone\otimes\sigma_z) |\Phi^+\rangle &=& |\Phi^-\rangle \nonumber\\
(\openone\otimes\sigma_x) |\Phi^+\rangle &=& |\Psi^+\rangle \nonumber\\
(\openone\otimes\sigma_x\sigma_z) |\Phi^+\rangle &=& |\Psi^-\rangle
\end{eqnarray}
Therefore, if the input qubit $X$ of the Pauli channel is
maximally entangled with a reference qubit $R$ that is unchanged
while $X$ is processed by the channel, say if their joint state is
\begin{equation}
|\psi\rangle_{RX} = |\Phi^+\rangle \; ,
\end{equation} 
then the joint state of $R$ and the output $Y$
is a mixture of the four Bell states
\begin{equation}   \label{eq_mixture-Bell}
\rho_{RY} = 
 (1-p)\, |\Phi^+\rangle\langle \Phi^+|
+p_z \, |\Phi^-\rangle\langle \Phi^-|
+p_x \, |\Psi^+\rangle\langle \Psi^+|
+p_y \, |\Psi^-\rangle\langle \Psi^-|  \; ,
\end{equation}
with $p=p_x+p_y+p_z$. This Bell mixture uniquely characterizes
the Pauli channel since the weights of the Bell states
are simply the probabilities
associated with the four error operators. This alternate description
of a Pauli channel based on Bell mixtures happens 
to be very useful when considering quantum cloning machines.
\par

A simple correspondence rule can be written
relating an arbitrary mixture of Bell state and the associated
operation on a qubit $|\psi\rangle$ by a Pauli channel.
Start from the Bell mixture 
\begin{equation}
\rho_{RY} = (1-p)\, |\Phi^+\rangle\langle \Phi^+|
+ \sum_{i=1}^3 p_i |\Psi_i\rangle\langle \Psi_i|
\end{equation}
where $p_1\le p_2\le p_3$, $p=p_1+p_2+p_3$, and $|\Psi_i\rangle$ stand for 
the three remaining Bell states ranked by increasing weight. 
It is straightforward to show that
the operation on an arbitrary state $|\psi\rangle$ performed
by the corresponding channel is
\begin{equation}   \label{eq_arbitrarychannel}
|\psi\rangle \to \rho=(1-p-p_2) \; |\psi\rangle\langle\psi|
+ (p_2-p_1) \; \sigma_1|\psi_{\perp}\rangle\langle\psi_{\perp}|\sigma_1
+ (p_3-p_2) \; \sigma_3|\psi\rangle\langle\psi|\sigma_3
+ 2(p_1+p_2) \; \openone/2
\end{equation}
where $|\psi_{\perp}\rangle=-i\sigma_y|\psi^*\rangle
=\sigma_x\sigma_z|\psi^*\rangle$ denotes
the time-reversed of state $|\psi\rangle$.
The four components on the right-hand side of Eq.~(\ref{eq_arbitrarychannel})
correspond respectively to the unchanged, (rotated) time-reversed,
rotated, and random fractions. It is clear from
Eq.~(\ref{eq_arbitrarychannel}) that the operation of the channel
is {\em state-independent} only if $p_1=p_2=p_3=p/3$,
that is, if the time-reversed and rotated fractions
vanish. Then, we have a {\em depolarizing} channel of probability $p$,
i. e., $\rho_{RX}$ is a Werner state 
and Eq.~(\ref{eq_arbitrarychannel}) becomes
\begin{equation}
|\psi\rangle \to \rho=(1-4p/3) \; |\psi\rangle\langle\psi|
+ (4p/3) \; \openone/2
\end{equation}
Thus, the qubit is replaced by a random bit with probability
$\pi=4p/3$ and left unchanged otherwise. The quantity $\pi$
is named the {\em depolarizing fraction}.
Equivalently, the vector characterizing the input qubit in the Bloch sphere
is shrunk by a {\em scaling factor} $s=1-4p/3$ regardless its orientation,
so that the fidelity of the channel,
$f = \langle\psi|\rho|\psi\rangle = 1-2p/3=(1+s)/2$,
is independent of the input state.
Other channels are necessarily {\em state-dependent}. For example,
the ``2-Pauli'' channel of probability $p$
(i.e., $p_x=p_z=p/2$ and $p_y=0$) performs the operation
\begin{eqnarray}  \label{eq_2Paulichannel}
|\psi\rangle \to \rho &=& (1-3p/2) \; |\psi\rangle\langle\psi|
+ (p/2) \; \sigma_y|\psi_{\perp}\rangle\langle\psi_{\perp}|\sigma_y
+ p \; \openone/2 \nonumber \\
&=& (1-3p/2) \; |\psi\rangle\langle\psi|
+ (p/2) \; |\psi^*\rangle\langle\psi^*|
+ p \; \openone/2
\end{eqnarray}
while the dephasing channel of probability $p$
(i.e., $p_z= p$ and $p_x=p_y=0$) simply gives
\begin{equation}
|\psi\rangle \to (1-p) \; |\psi\rangle\langle\psi|
+ p \; \sigma_z|\psi\rangle\langle\psi|\sigma_z
\end{equation}

\subsection{Asymmetric Pauli cloning machines}
\label{sect_PCM}

\begin{figure}
\caption{Pauli cloning machine of input $X$ (initially entangled
with a reference $R$) and outputs $A$ and
$B$. The third output $C$ refers to an ancilla or 
the cloning machine. The three outputs emerge in general
from distinct Pauli channels.}
\vskip 0.25cm
\centerline{\psfig{file=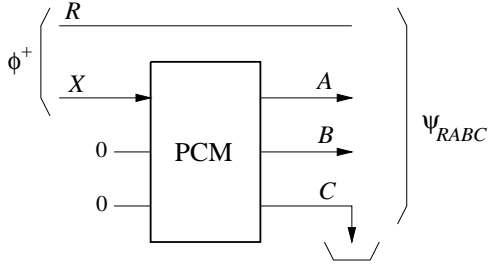,width=2.5in,angle=-90}}
\label{fig_cloner}
\vskip -0.25cm
\end{figure}

We define an {\em asymmetric} Pauli cloning machine
as a machine whose two outputs, $A$ and $B$,
emerge from distinct Pauli channels~\cite{bib_cerf}.
Thus, if the input $X$ of the cloner is fully entangled
with a reference $R$, i.e., $|\psi\rangle_{RX}=|\Phi^+\rangle$,
the density operators $\rho_{RA}$
and $\rho_{RB}$ must then be mixtures of Bell states.
Focusing on the first output $A$, we see that
a 4-dimensional additional Hilbert space is necessary in general
to ``purify'' $\rho_{RA}$ since we need to accommodate its four
(generally nonzero) eigenvalues.\footnote{In other words, if
$\rho_{RA}$ results from the partial trace of a pure state
in an extended Hilbert space, this space must be 16-dimensional
as a consequence of Schmidt decomposition~\cite{bib_schmidt}.} 
The 2-dimensional space of the second output qubit $B$ is thus 
insufficient for this purpose,
so that we must introduce a single additional qubit $C$,
which may be viewed as an ancilla or the cloning machine itself.
The fact that a 2-dimensional space is sufficient for $C$ is
shown in Refs.~\cite{bib_bh,bib_bruss} for the UCM, and we conjecture
that this holds for the PCM (this is justified a posteriori
by the work of Niu and Griffiths~\cite{bib_niu}).
As a consequence, we are led to consider a 4-qubit system 
in order to fully describe the PCM, as pictured in Fig.~\ref{fig_cloner}.
The qubits $R$ and $X$ are initially in the entangled state 
$|\Phi^+\rangle$, the two auxiliary qubits being in a prescribed
state, e.g., $|0\rangle$. After cloning, 
the four qubits $R$, $A$, $B$, and $C$
are in a pure state for which $\rho_{RA}$ and $\rho_{RB}$ are
mixtures of Bell states (i.e., $A$ and $B$ emerge from a Pauli channel).
As we shall see, $\rho_{RC}$ happens to be also a Bell mixture, so that
$C$ can be viewed as a third output emerging from another Pauli channel.
\par

Instead of specifying a Pauli cloning machine by a particular
unitary transformation acting on the input state $|\psi\rangle$
(together with the two auxiliary qubits in state $|0\rangle$),
we choose here to characterize a PCM by the
wave function $|\Psi\rangle_{RABC}$ underlying the
entanglement of the three outputs with $R$.
The question is thus to find in general the 4-qubit wave functions
that satisfy the requirement that the state of every qubit pair is
a mixture of the four Bell states. (It appears that, if $RA$ and $RB$
are Bell mixtures, so are all pairs among $RABC$.)
Making use of the Schmidt decomposition~\cite{bib_schmidt}
of $|\Psi\rangle_{RABC}$ for the bipartite partition $RA$ vs $BC$, 
we see that this state can be written
as a superposition of {\em double Bell} states
\begin{equation}  \label{eq_psi}
|\Psi\rangle_{RA;BC}= 
\left\{ v \, |\Phi^+\rangle |\Phi^+\rangle 
+z \, |\Phi^-\rangle |\Phi^-\rangle 
+x \, |\Psi^+\rangle |\Psi^+\rangle 
+y \, |\Psi^-\rangle |\Psi^-\rangle  \right\}_{RA;BC}  \; ,
\end{equation}
where $x$, $y$, $z$, and $v$ are complex amplitudes
(with $|x|^2+|y|^2+|z|^2+|v|^2=1$).
Note that the possible permutations of the Bell states in Eq.~(\ref{eq_psi})
are not considered here for simplicity.
The requirement that the qubit pairs $RA$ and $BC$ are Bell mixtures
is thus satisfied, that is, $\rho_{RA}=\rho_{BC}$ is of the form
of Eq.~(\ref{eq_mixture-Bell})
with $p_x=|x|^2$, $p_y=|y|^2$, $p_z=|z|^2$, and $1-p=|v|^2$.
A remarkable feature of these double Bell states 
is that they transform into superpositions of double Bell states 
for the two other possible partitions of the four qubits $RABC$ 
into two pairs ($RB$ vs $AC$, $RC$ vs $AB$). For example, 
the transformation associated with the partition $RB$ vs $AC$ is
\begin{eqnarray}
|\Phi^+\rangle_{RA} \; |\Phi^+\rangle_{BC} &=&
{1\over 2} \left\{ |\Phi^+\rangle |\Phi^+\rangle 
+|\Phi^-\rangle |\Phi^-\rangle 
+|\Psi^+\rangle |\Psi^+\rangle 
+|\Psi^-\rangle |\Psi^-\rangle  \right\}_{RB;AC}  \nonumber \\
|\Phi^-\rangle_{RA} \; |\Phi^-\rangle_{BC} &=&
{1\over 2} \left\{ |\Phi^+\rangle |\Phi^+\rangle 
+|\Phi^-\rangle |\Phi^-\rangle 
-|\Psi^+\rangle |\Psi^+\rangle 
-|\Psi^-\rangle |\Psi^-\rangle  \right\}_{RB;AC}  \nonumber \\
|\Psi^+\rangle_{RA} \; |\Psi^+\rangle_{BC} &=&
{1\over 2} \left\{ |\Phi^+\rangle |\Phi^+\rangle 
-|\Phi^-\rangle |\Phi^-\rangle 
+|\Psi^+\rangle |\Psi^+\rangle 
-|\Psi^-\rangle |\Psi^-\rangle  \right\}_{RB;AC}  \nonumber \\
|\Psi^-\rangle_{RA} \; |\Psi^-\rangle_{BC} &=&
{1\over 2} \left\{ |\Phi^+\rangle |\Phi^+\rangle 
-|\Phi^-\rangle |\Phi^-\rangle 
-|\Psi^+\rangle |\Psi^+\rangle 
+|\Psi^-\rangle |\Psi^-\rangle  \right\}_{RB;AC}
\end{eqnarray}
(For the partition $RC$ vs $AB$, these expressions are similar
up to an overall sign in the transformation of the
state $|\Psi^-\rangle_{RA} \; |\Psi^-\rangle_{BC}$.)
This implies that $|\Psi\rangle_{RABC}$ is also a superposition
of double Bell states (albeit with different amplitudes) for
these two other partitions, which, therefore, also yield mixtures
of Bell states when tracing over half of the system.
Specifically, for the partition $RB$ vs $AC$, we obtain
\begin{equation}
|\Psi\rangle_{RB;AC}= 
\left\{ v' \, |\Phi^+\rangle |\Phi^+\rangle 
+z' \, |\Phi^-\rangle |\Phi^-\rangle 
+x' \, |\Psi^+\rangle |\Psi^+\rangle 
+y' \, |\Psi^-\rangle |\Psi^-\rangle  \right\}_{RB;AC}  \; ,
\end{equation}
with
\begin{eqnarray}  \label{eq_vzxy'}
v' &=& (v+z+x+y)/2 \nonumber\\
z' &=& (v+z-x-y)/2 \nonumber\\
x' &=& (v-z+x-y)/2 \nonumber\\
y' &=& (v-z-x+y)/2 
\end{eqnarray}
implying that the second output $B$ emerges from a Pauli
channel with probabilities $p_x'=|x'|^2$, $p_y'=|y'|^2$, and $p_z'=|z'|^2$.
Similarly, the third output $C$ is described by considering 
the partition $RC$ vs $AB$,
\begin{equation}
|\Psi\rangle_{RC;AB}= 
\left\{ v'' \, |\Phi^+\rangle |\Phi^+\rangle 
+z'' \, |\Phi^-\rangle |\Phi^-\rangle 
+x'' \, |\Psi^+\rangle |\Psi^+\rangle 
+y'' \, |\Psi^-\rangle |\Psi^-\rangle  \right\}_{RC;AB}  \; ,
\end{equation}
with
\begin{eqnarray}   \label{eq_vzxy''}
v'' &=& (v+z+x-y)/2  \nonumber\\
z'' &=& (v+z-x+y)/2  \nonumber\\
x'' &=& (v-z+x+y)/2  \nonumber\\
y'' &=& (v-z-x-y)/2
\end{eqnarray}
Thus, Eqs.~(\ref{eq_vzxy'}) and (\ref{eq_vzxy''}) 
relate the amplitudes of the double Bell states for the three possible
partitions of the four qubits into two pairs, and thereby
specify the entire set of asymmetric Pauli cloning machines.
\par

\subsection{No-cloning inequality for quantum bits}

The complementarity between the two copies
produced by an asymmetric PCM can be shown to result in general
from an uncertainty principle, much like that
associated with Fourier transforms. In order to show this, let
us rewrite the amplitudes of $|\psi\rangle_{RA;BC}$ as a
two-dimensional discrete function $\alpha_{m,n}$ with $m,n=0,1$:
\begin{eqnarray}
\alpha_{0,0}&=&v \nonumber\\
\alpha_{0,1}&=&z \nonumber\\
\alpha_{1,0}&=&x \nonumber\\
\alpha_{1,1}&=&y
\end{eqnarray}
Thus, output $A$ emerges from a Pauli channel characterized
by the probability distribution $p_{m,n}=|\alpha_{m,n}|^2$,
where $p_{0,1}=p_z$, $p_{1,0}=p_x$, $p_{1,1}=p_y$, and $p_{0,0}$ is
simply the probability that the qubit remains unchanged.
Similarly, output $B$ can be characterized by a
two-dimensional function $\beta_{m,n}$ defined as
\begin{eqnarray}
\beta_{0,0}&=&v' \nonumber\\
\beta_{0,1}&=&z' \nonumber\\
\beta_{1,0}&=&x' \nonumber\\
\beta_{1,1}&=&y'
\end{eqnarray}
resulting in the probability distribution $p_{m,n}'=|\beta_{m,n}|^2$.
Using this notation, it appears that Eq.~(\ref{eq_vzxy'}) is simply
a two-dimensional discrete Fourier transform,\footnote{This result
will be shown to hold in $N$ dimensions in Sec.~\ref{sect_Ndim}.}
\begin{equation}   \label{eq_DFT-qubits}
\beta_{m,n}={1\over 2} \sum_{x=0}^1 \sum_{y=0}^1 (-1)^{nx+my} \; \alpha_{x,y}
\end{equation}
This emphasizes that if output $A$ is close to perfect ($\alpha_{m,n}$ is
a peaked function) then output $B$ is very noisy ($\beta_{m,n}$ is
a flat function), and conversely. 
Consequently, the probability distributions $p_{m,n}=|\alpha_{m,n}|^2$
and $p_{m,n}'=|\beta_{m,n}|^2$ characterizing
the channels leading to outputs $A$ and $B$ cannot 
have a variance simultaneously tending to zero,
giving rise to an uncertainty principle that governs
the tradeoff between the quality of the copies.
\par

To illustrate this, let us consider the class 
of {\em isotropic} asymmetric PCMs, i.e., cloners
whose outputs $A$ and $B$ emerge from (distinct) 
{\em depolarizing} channels. As we will show later on,
the corresponding set of conditions
\begin{eqnarray}  \label{eq_2isotropconditions}
|x|&=&|y|=|z| \nonumber\\
|x'|&=&|y'|=|z'|
\end{eqnarray}
simply implies
\begin{eqnarray}  \label{eq_2isotrop}
x&=&y=z \nonumber\\
x'&=&y'=z'
\end{eqnarray}
Consider a PCM whose output $A$ emerges 
from a depolarizing channel of probability $p=3|x|^2$, i.e.,
\begin{equation}  \label{eq_asymm1}
\rho_{RA}=
 |v|^2 |\Phi^+\rangle\langle \Phi^+|
+|x|^2 \left( |\Phi^-\rangle\langle \Phi^-|
+|\Psi^+\rangle\langle \Psi^+|
+|\Psi^-\rangle\langle \Psi^-|  \right)  \; ,
\end{equation}
with $|v|^2+3|x|^2=1$. Then, from Eq.~(\ref{eq_vzxy'}),
we have $v'=(v+3x)/2$ and $x'=y'=z'=(v-x)/2$, resulting in
\begin{equation}   \label{eq_asymm2}
\rho_{RB}= {|v+3x|^2 \over 4}
|\Phi^+\rangle\langle \Phi^+|
+{|v-x|^2 \over 4} \left( |\Phi^-\rangle\langle \Phi^-|
+ |\Psi^+\rangle\langle \Psi^+|
+ |\Psi^-\rangle\langle \Psi^-| \right)  \; .
\end{equation}
Thus, the second output $B$ also emerges from a depolarizing channel
of probability $p'=3|x'|^2={3\over 4}|v-x|^2$.
In other words, both outputs of this PCM are state-independent 
as the vector characterizing the input state $|\psi\rangle$ 
in the Bloch sphere undergoes a (different) shrinking at each
output, regardless of its orientation. 
Note that the third output $C$ emerges in general from 
a different (state-dependent) Pauli channel. Using the normalization
condition, the relation between the parameters $x$ and $x'$
characterizing the two outputs can be written as
\begin{equation}  \label{eq_cross}
|x|^2+{\rm Re}(x^* x')+|x'|^2= {1\over 4}
\end{equation}
that is, an ellipse representing a set of cloners in the
$(x,x')$-space. By varying the
relative phase of $x$ and $x'$ (the global phase is irrelevant),
one varies the eccentricity of this ellipse.
Clearly, the best cloning (minimum values for $|x|$ and $|x'|$)
is achieved when the cross term in Eq.~(\ref{eq_cross})
is the largest in magnitude,
that is when $x$ and $x'$ have the same (or opposite) phases. 
We may thus assume that $x$ and $x'$ are real and positive without
loss of generality.
Consequently, the tradeoff between the quality of the copies
can be described by the {\em no-cloning inequality}
\begin{equation}  \label{eq_no-cloning-ineq}
x^2+xx'+x'^2 \ge {1\over 4} \; ,
\end{equation}
where the copying error is measured by the probability of the
depolarizing channel underlying each output, i.e., 
$p=3x^2$ and $p'=3x'^2$ (with $x,x'\ge 0$).
Provided that a single additional qubit $C$ is sufficient for the cloner,
the imperfect cloning achieved by such an isotropic PCM 
is {\em optimal}: the PCM achieves the minimum $p$ and $p'$ 
for a fixed ratio $p/p'$.
Thus, Eq.~(\ref{eq_no-cloning-ineq})
is the tightest no-cloning bound that can be written 
for a qubit.\footnote{Equation (\ref{eq_no-cloning-ineq})
has been independently derived by Niu and Griffiths~\cite{bib_niu}.}
\par

\begin{figure}
\caption{Ellipse delimiting the best quality of the two outputs of
an asymmetric PCM that can be achieved simultaneously
(only the quadrant $x,x'\ge 0$ is of interest here). The outputs
emerge from depolarizing channels of probability $p=3x^2$ 
and $p'=3x'^2$. Any close-to-perfect cloning characterized 
by a point inside the ellipse is forbidden.}
\vskip 0.25cm
\centerline{\psfig{file=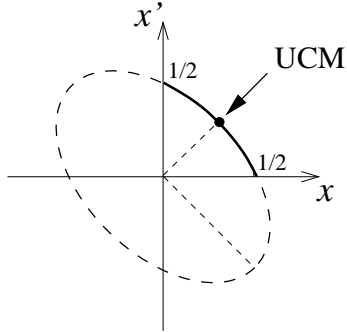,width=1.8in,angle=-90}}
\label{fig_nocloning}
\vskip -0.25cm
\end{figure}

Equation~(\ref{eq_no-cloning-ineq}) corresponds to the domain
in the $(x,x')$-space located outside an ellipse
whose semiminor axis, oriented 
in the direction $(1,1)$, is $1/\sqrt{6}$,
as shown in Fig.~\ref{fig_nocloning}.
(The semimajor axis is $1/\sqrt{2}$.) 
The origin in this space corresponds to a (nonexisting) cloner whose
two outputs would be perfect $p=p'=0$, while to distance
to origin measures $(p+p')/3$. The ellipse characterizes
the ensemble of values for $p$ and $p'$ that can actually be achieved
with an optimal PCM. It intercepts its minor axis
at $(1/\sqrt{12},1/\sqrt{12})$, which corresponds to the
universal cloning machine (UCM), i.e., $p=p'=1/4$.
This point is the closest to the origin 
(i.e., the cloner with minimum $p+p'$), 
and characterizes in this sense the best possible
copying.\footnote{The fact that 
the UCM ($p=p'$) is optimal was proven 
in Refs.~\cite{bib_bruss,bib_gisinmassar}.}
The UCM is the only symmetric cloner belonging to the
class of isotropic PCMs considered here (i.e., cloners 
whose outputs are depolarizing channels); 
other symmetric---but anisotropic---cloners will be 
considered in Sec.~\ref{sect_symmPCM}.
The ellipse crosses the $x$-axis at $(1/2,0)$, which describes the
situation where the first output emerges from a 100\%-depolarizing 
channel ($p=3/4$) while the second emerges from a perfect
channel ($p'=0$). Of course, $(0,1/2)$ corresponds to the symmetric
situation. 
\par 

Introducing a phase difference between $x$ and $x'$
results in a set of PCMs characterized by
an ellipse that is less eccentric and tends to a circle
of radius $1/2$ for a phase difference
of $\pi/2$. Consequently, the no-cloning inequality~(\ref{eq_no-cloning-ineq})
is saturated when $x$ and $x'$ have the same (or opposite) phase.
The domain inside the ellipse corresponds therefore 
to the values for $p$ and $p'$
that cannot be achieved simultaneously, reflecting the
impossibility of close-to-perfect cloning.
\par

Let us show that
the no-cloning inequality can be recast in terms of the depolarizing
fractions $\pi_a$ and $\pi_b$ underlying outputs $A$ and $B$.
As noted before, Eq.~(\ref{eq_vzxy'})
can be written as a unitary transformation $U$ (a 2-D discrete Fourier
transform):
\begin{equation}  \label{eq_eigendecompos}
\left( \begin{array}{c} v' \\ z' \\ x' \\ y' \end{array} \right)
=  \underbrace { {1\over 2} \left( \begin{array}{r r r r}
1 & 1 & 1 & 1 \\
1 & 1 &-1 &-1 \\
1 &-1 & 1 &-1 \\
1 &-1 &-1 & 1
\end{array} \right) }_U
\;
\left( \begin{array}{c} v \\ z \\ x \\ y \end{array} \right)
\end{equation}
This transformation admits the (three-fold degenerate) 
eigenvalue $\lambda_1=1$,
associated with the eigenspace $v=x+y+z$, and eigenvalue $\lambda_2=-1$,
associated with the eigenvector $x=y=z=-v$.
Then, for the conditions (\ref{eq_2isotropconditions})
to hold simultaneously, we must have
\begin{equation} \label{eq_c+d}
\left( \begin{array}{c} v \\ z \\ x \\ y \end{array} \right)
= c \left( \begin{array}{r} 3 \\ 1 \\ 1 \\ 1 \end{array} \right)
+ d \left( \begin{array}{r} -1 \\ 1 \\ 1 \\ 1 \end{array} \right)
\end{equation}
and
\begin{equation}  \label{eq_c-d}
\left( \begin{array}{c} v' \\ z' \\ x' \\ y' \end{array} \right)
= c \left( \begin{array}{r} 3 \\ 1 \\ 1 \\ 1 \end{array} \right)
- d \left( \begin{array}{r} -1 \\ 1 \\ 1 \\ 1 \end{array} \right)
\end{equation}
where $c$ and $d$ are complex numbers
with normalization $12|c|^2+4|d|^2=1$.
Note that this implies Eq.~(\ref{eq_2isotrop}), as claimed before.
Equation~(\ref{eq_c+d}) can be rewritten as
\begin{equation}
\left( \begin{array}{c} v \\ z \\ x \\ y \end{array} \right)
= \hat{a} \left( \begin{array}{r} 1 \\ 0 \\ 0 \\ 0 \end{array} \right)
+ {a\over 2} 
  \left( \begin{array}{r} 1 \\ 1 \\ 1 \\ 1 \end{array} \right)
\end{equation}
where $a=2(c+d)$ and $\hat{a}=2(c-d)$ correspond to the
``flat'' and ``peaked'' components of the first output, respectively. 
Similarly, for the second output, we can rewrite Eq.~(\ref{eq_c-d}) as
\begin{equation}
\left( \begin{array}{c} v' \\ z' \\ x' \\ y' \end{array} \right)
= \hat{b} \left( \begin{array}{r} 1 \\ 0 \\ 0 \\ 0 \end{array} \right)
+ {b\over 2} 
  \left( \begin{array}{r} 1 \\ 1 \\ 1 \\ 1 \end{array} \right)
\end{equation}
where $b=2(c-d)$ and $\hat{b}=2(c+d)$ correspond again to the
``flat'' and ``peaked'' components.
Using $c=(a+b)/4$, $d=(a-b)/4$, and the normalization condition,
we obtain
\begin{equation}
|a|^2+{\rm Re}(a^* b) + |b|^2 =1
\end{equation}
For the same reason as before, it is sufficient to consider
$a$ and $b$ real and positive for an optimal cloner. Therefore,
an alternate no-cloning inequality can be expressed as
\begin{equation}  \label{eq_no_clon_uncert}
a^2+ab+b^2 \ge 1 \; ,
\end{equation}
where $\pi_a=a^2=4x^2=4p/3$ and $\pi_b=b^2=4x'^2=4p'/3$ are 
the {\em depolarizing fraction} of the channels leading
to outputs $A$ and $B$, 
respectively. We have indeed
\begin{eqnarray}
\rho_{RA}&=&(1-a^2) \; |\Phi^+\rangle\langle\Phi^+| 
+a^2 \; {\openone\otimes\openone \over 4} \nonumber\\
\rho_{RB}&=&(1-b^2) \; |\Phi^+\rangle\langle\Phi^+| 
+b^2 \; {\openone\otimes\openone \over 4}
\end{eqnarray}
so that the input qubit is replaced by a random qubit 
with probability $\pi_a$ ($\pi_b$) or left unchanged with probability
$1-\pi_a$ ($1-\pi_b$) in the channel leading to output $A$ ($B$).
Equation~(\ref{eq_no_clon_uncert}) is equivalent to
Eq.~(\ref{eq_no-cloning-ineq}) by substituting $x=a/2$ and $x'=b/2$.
The ellipse corresponding to Eq.~(\ref{eq_no_clon_uncert})
crosses the axes at $(0,1)$ and $(1,0)$,
while the closest point to the origin is $(\sqrt{1/3},\sqrt{1/3})$
and coincides with the UCM.
Indeed, the outputs of the UCM emerge from two channels whose
depolarizing fraction is $\pi_a=\pi_b=1/3$. 
The no-cloning bound associated with $N$-dimensional
quantum states (instead of qubits)
will be investigated in Sec.~\ref{sect_Ndim_nocloning}.
We will see that the cross-term in Eq.~(\ref{eq_no_clon_uncert})
is replaced by $2ab/N$, implying that the ellipse
tends to a circle of radius one at the limit of a large dimension $N$.
\par

\subsection{No-cloning uncertainty relation}

The complementarity between the index $m$ of $\alpha_{m,n}$ and the index
$n$ of $\beta_{m,n}$ (or conversely) implied by Eq.~(\ref{eq_DFT-qubits})
can be expressed in a more general way by using the uncertainty relation
(or Robertson relation)
\begin{equation}  \label{eq_robertson}
\langle \Delta O_A^2 \rangle \; \langle \Delta O_B^2 \rangle
\ge {1\over 4} |\langle [O_A,O_B] \rangle|^2
\end{equation}
where $O_A$ and $O_B$ are two observables, while 
$\Delta O_A = O_A-\langle O_A \rangle$, and
$\Delta O_B = O_B-\langle O_B \rangle$. This relation holds when the
quantum expectation values are taken for any quantum state.
Consider the state
$|\psi\rangle=v|00\rangle+z|01\rangle+x|10\rangle+y|11\rangle$, and
choose $O_A=\sigma_z/2\otimes \openone$ and $O_B=\sigma_x/2\otimes
\openone$, so that $[O_A,O_B]=i\sigma_y/2$.
Applying Eq.~(\ref{eq_robertson}) to the state $|\psi\rangle$ yields
the no-cloning uncertainty relation
\begin{equation}  \label{eq_robertson_cloning}
(\underbrace{|v|^2+|z|^2}_{1-p_x-p_y})
(\underbrace{|x|^2+|y|^2}_{p_x+p_y}) \times
(\underbrace{|v'|^2+|x'|^2}_{1-p_z'-p_y'})
(\underbrace{|z'|^2+|y'|^2}_{p_z'+p_y'})
\ge {1\over 4} | {\rm Im}(v^*x+z^*y) |^2
\end{equation}
The term $(1-p_x-p_y)(p_x+p_y)$ is simply the
variance of the distribution $p_m=\sum_n p_{m,n}$ associated with
the first output,
while $(1-p_z'-p_y')(p_z'+p_y')$ is the
variance of $p_n'=\sum_m p_{m,n}'$ (associated with the second output).
Equation~(\ref{eq_robertson_cloning}) gives thus a lower bound on
the product of the variances of $p_m$ and $p_n'$.
(This bound depends on the state $|\psi\rangle$.) 
It is easy to check that
when the distribution $p_m$ is peaked ($x=y=0$ or $v=z=0$)
the lower bound tends to zero, as expected.
The bound can also be reexpressed as
\begin{equation}
{1\over 4} | {\rm Im}(v'^*z'+x'^*y') |^2
\end{equation}
implying that it also tends to zero when $p_n'$ is peaked ($z'=y'=0$
or $v'=x'=0$). Unfortunately, the bound is not saturated
in the case of the UCM. Using Eq.~(\ref{eq_robertson}) with
$|\psi'\rangle=v'|00\rangle+z'|01\rangle+x'|10\rangle+y'|11\rangle$
one obtains a alternate inequality expressing the
duality between $p_n=\sum_m p_{m,n}$ and $p_m'=\sum_n p_{m,n}'$,
\begin{equation}  \label{eq_robertson_cloning2}
(\underbrace{|v|^2+|x|^2}_{1-p_z-p_y})
(\underbrace{|z|^2+|y|^2}_{p_z+p_y}) \times
(\underbrace{|v'|^2+|z'|^2}_{1-p_x'-p_y'})
(\underbrace{|x'|^2+|y'|^2}_{p_x'+p_y'})
\ge {1\over 4} | {\rm Im}(v^*z+x^*y) |^2
\end{equation}
In Sect.~\ref{sect_entropicnocloning}, 
more general no-cloning uncertainty relations
will be derived, based on the {\em entropic} uncertainty relations
for non-commuting observables.

\subsection{Symmetric Pauli cloning machines}
\label{sect_symmPCM}

\begin{figure}
\caption{Oblate ellipsoid representing the class of symmetric PCMs
whose two outputs emerge from the same Pauli channel
of parameters $p_x=x^2$, $p_y=y^2$, and $p_z=z^2$
(only the octant $x,y,z\ge 0$ is considered here). The pole of this
ellipsoid corresponds to the UCM.
The capacity of a Pauli channel that lies outside this ellipsoid
must be vanishing.}
\vskip 0.25cm
\centerline{\psfig{file=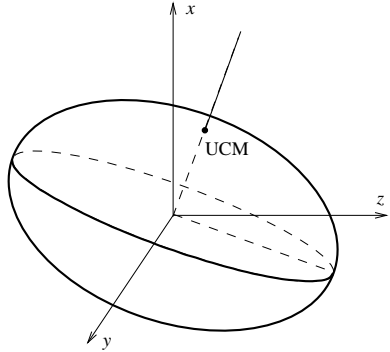,width=2.0in,angle=-90}}
\label{fig_ellipsoid}
\vskip -0.25cm
\end{figure}

It was shown in Ref.~\cite{bib_bruss} that
an interesting application of the UCM is that 
it can be used to establish an upper bound on the quantum capacity $C$
of a depolarizing channel, namely $C=0$ at $p=1/4$. This result
relies on the fact that the UCM is symmetric, as we will see
in Sec.~\ref{sect_bound}. It is therefore natural to extend
this to the Pauli cloning machines.
Consider the class of symmetric PCMs that have both outputs
emerging from a {\em same} (but not necessarily isotropic)
Pauli channel, i.e., $\rho_{RA}=\rho_{RB}$. Thus,
these PCMs must satisfy the conditions
\begin{eqnarray}
|v'|&=&|v| \nonumber \\
|z'|&=&|z| \nonumber \\
|x'|&=&|x| \nonumber \\
|y'|&=&|y|
\end{eqnarray}
The eigenspectrum decomposition of the operator $U$ 
[cf. Eq.~(\ref{eq_eigendecompos})] implies that these conditions
hold for any vector in the eigenspace associated with $\lambda_1=1$,
or for the eigenvector $x=y=z=-v=1/2$ associated with $\lambda_2=-1$.
The latter solution corresponds to a trivial PCM whose two
outputs are fully depolarizing. The interesting solution is thus
\begin{equation}  \label{eq_condition-dupl}
v=x+y+z  \; ,
\end{equation}
where $x$, $y$, $z$, and $v$ can be assumed to be real.
Equation~(\ref{eq_condition-dupl}), together with the normalization
condition, describes a two-dimensional surface in a space
where each point $(x,y,z)$ represents a Pauli channel of parameters
$p_x=x^2$, $p_y=y^2$, and $p_z=z^2$ (We only consider here the first
octant $x,y,z\ge 0$). This surface,
\begin{equation}  \label{eq_ellipsoid}
x^2+y^2+z^2+xy+xz+yz={1\over 2}  \; ,
\end{equation}
is an oblate ellipsoid $E$ with symmetry axis
along the direction $(1,1,1)$, as shown in Fig.~\ref{fig_ellipsoid}.
The semiminor axis (or polar radius)
is $1/2$ while the semimajor axis (or equatorial radius) is $1$. 
In this representation, the distance to the origin 
is $p_x+p_y+p_z$,
so that the pole $(1/\sqrt{12},1/\sqrt{12},1/\sqrt{12})$
of this ellipsoid---the closest point to the origin---corresponds
to the special case of a depolarizing channel
of probability $p=1/4$. Thus, this particular symmetric
PCM reduces to the UCM and is isotropic. 
This simply illustrates that
the requirement of having an optimal cloning (minimum $p_x+p_y+p_z$)
implies that the cloner is state-independent ($p_x=p_y=p_z$).
The parametric equations of ellipsoid $E$ are
\begin{eqnarray}
x&=&\sqrt{1\over 12}\,\cos(\theta) + \sqrt{2\over 3}\,\sin(\theta)\cos(\phi)
\nonumber\\
y&=&\sqrt{1\over 12}\,\cos(\theta) + \sqrt{2\over 3}\,\sin(\theta)
\cos(\phi+2\pi/3)
\nonumber\\
z&=&\sqrt{1\over 12}\,\cos(\theta) + \sqrt{2\over 3}\,\sin(\theta)
\cos(\phi+4\pi/3)
\end{eqnarray}
where the polar angle $\theta$ measures the ``distance'' from the depolarizing
channel underlying the UCM ($\theta=0$ implies $p_x=p_y=p_z$), while the 
azimuthal angle $\phi$
characterizes the distribution among $p_x$, $p_y$, and $p_z$.
\par
\par

\subsection{Universal cloning machine}

The optimal symmetric PCM (i.e., the UCM) can be obtained alternatively
by requiring that the two outputs $A$ and $B$ of a symmetric cloner are
maximally independent.
Using Eqs.~(\ref{eq_vzxy''}) and (\ref{eq_condition-dupl}),
we obtain 
\begin{eqnarray}
v''&=&x+z \nonumber\\ 
z''&=&y+z \nonumber\\
x''&=&x+y \nonumber\\
y''&=&0
\end{eqnarray}
Therefore, we have
\begin{equation}   \label{eq_asymm3}
\rho_{RC}= \rho_{AB}= |x+z|^2\; |\Phi^+\rangle\langle \Phi^+|
+ |y+z|^2 \; |\Phi^-\rangle\langle \Phi^-|
+ |x+y|^2 \; |\Psi^+\rangle\langle \Psi^+| \; .
\end{equation}
(This means that the third output $C$ emerges from 
a Pauli channel with vanishing $p_y$.)
Thus, we need to maximize the joint von Neumann entropy
of the two outputs $A$ and $B$,
\begin{equation}
S(AB)=-{\rm Tr}(\rho_{AB} \log\rho_{AB} ) =
H \left[ |x+z|^2,|y+z|^2,|x+y|^2 \right]
\end{equation}
with $H[\cdot]$ denoting the Shannon entropy.
It is easy to see that the solution with $x,y,z\ge 0$
that maximizes $S(AB)$ is $x=y=z$, that is,
the Pauli channel underlying outputs $A$ and $B$
reduces to a depolarizing channel.
Using Eq.~(\ref{eq_ellipsoid}), we get $x=y=z=1/\sqrt{12}$, so that
the wave function underlying the UCM is
\begin{equation}  \label{eq_psi-UCM}
|\Psi\rangle_{RA;BC}
=\sqrt{3\over 4} \; |\Phi^+\rangle_{RA} |\Phi^+\rangle_{BC}
+\sqrt{1\over12} \left\{
|\Phi^-\rangle|\Phi^-\rangle
+|\Psi^+\rangle|\Psi^+\rangle
+|\Psi^-\rangle|\Psi^-\rangle \right\}_{RA;BC}  \\
\end{equation}
Consequently
\begin{equation}
\rho_{RA}=\rho_{RB}= {3\over 4}\; |\Phi^+\rangle\langle\Phi^+|
+{1\over 12} \left( |\Phi^-\rangle\langle\Phi^-|
+ |\Psi^+\rangle\langle\Psi^+|
+ |\Psi^-\rangle\langle\Psi^-| \right)
\end{equation}
reflecting that $A$ and $B$ emerge both
from a depolarizing channel with $p=1/4$, 
\begin{equation}  \label{eq_UCMstandout}
|\psi\rangle \to {2\over 3} \; |\psi\rangle\langle\psi|
+ {1\over 3} \; (\openone/2)
\end{equation}
As mentioned above, the corresponding scaling factor is $s=1-4p/3=2/3$
while the fidelity of cloning is $f=1-2p/3=5/6$~\cite{bib_bh}.
For the partition $RC$ vs $AB$, we obtain
\begin{equation}  \label{eq_rcab}
|\Psi\rangle_{RC;AB} = \sqrt{1\over 3} \left\{
|\Phi^+\rangle |\Phi^+\rangle
+|\Phi^-\rangle|\Phi^-\rangle
+|\Psi^+\rangle|\Psi^+\rangle \right\}_{RC;AB}
\end{equation}
implying that the 4-qubit wave function is symmetric 
under the interchange of $A$ and $B$ (or $R$ and $C$).
It is easy to check that it corresponds to the unitary
transformation which implements the UCM~\cite{bib_bh}:
\begin{eqnarray} \label{eq_U_ucm}
|0\rangle_X \; |00\rangle &\to& \sqrt{2\over 3} |00\rangle_{AB} \; |0\rangle_C
+ \sqrt{1\over 3} |\Psi^+\rangle_{AB} \; |1\rangle_C  \nonumber\\
|1\rangle_X \; |00\rangle &\to& \sqrt{2\over 3} |11\rangle_{AB} \; |1\rangle_C
+ \sqrt{1\over 3} |\Psi^+\rangle_{AB} \; |0\rangle_C 
\end{eqnarray}
Indeed, using Eq.~(\ref{eq_U_ucm}), we have
\begin{equation}
|\Phi^+\rangle_{RX} |00\rangle \to 
\sqrt{1\over 3} \big( |00\rangle_{AB} \; |00\rangle_{RC} +
|11\rangle_{AB} \; |11\rangle_{RC} \big)
+ \sqrt{1\over 6} |\Psi^+\rangle_{AB} 
\big( |01\rangle_{RC}+|10\rangle_{RC} \big)
\end{equation}
if the initial state of $X$ is maximally entangled with the reference $R$.
Thus, the 4-qubit wave function
is transformed into Eq.~(\ref{eq_rcab}).
The latter implies
\begin{equation}
\rho_{RC} = \rho_{AB}= {1\over 3} \left(
|\Phi^+\rangle\langle\Phi^+|
+|\Phi^-\rangle\langle\Phi^-|
+|\Psi^+\rangle\langle\Psi^+| \right)
\end{equation}
showing that the joint entropy of the two outputs
is maximum (remember that the singlet $|\Psi^-\rangle$
component must vanish), or their mutual entropy is minimum.
This reflects that outputs $A$ and $B$
are maximally independent. Finally, we see also
that the third output of the UCM emerges from a 2-Pauli
channel of probability 2/3. Using Eq.~(\ref{eq_2Paulichannel}),
it appears that the corresponding operation on an arbitrary state
$|\psi\rangle$ is
\begin{equation}
|\psi\rangle \to {1\over 3} \; |\psi^*\rangle\langle\psi^*|
+ {2\over 3} \; (\openone/2)
\end{equation}
as noted in Ref.~\cite{bib_bh_registers}.

\subsection{Related bound on the capacity of the Pauli channel}
\label{sect_bound}

The class of symmetric PCMs 
characterized by Eq.~(\ref{eq_ellipsoid})
can be used in order to put a limit on the quantum capacity 
of a Pauli channel, thereby extending the result 
of Bruss et al.~\cite{bib_bruss} for the depolarizing channel.
Consider a PCM whose outputs emerge from a Pauli channel
of probabilities $p_x$, $p_y$, and $p_z$.
Applying an error-correcting scheme separately on each output
of the cloning machine (obliviously of the other output)
would lead to a violation of the no-cloning theorem if
the capacity $C(p_x,p_y,p_z)$ was nonzero. Since $C$ is a nonincreasing
function of $p_x$, $p_y$, and $p_z$, for $p_x,p_y,p_z\le 1/2$
(i.e., adding noise to a channel cannot increase its capacity),
we have
\begin{equation}
C(p_x,p_y,p_z)=0 \qquad {\rm if~} (x,y,z)\not\in E
\end{equation}
that is, the quantum capacity is vanishing for any Pauli channel
that lies {\em outside} the ellipsoid $E$. In particular,
Eq.~(\ref{eq_ellipsoid}) implies that the quantum capacity vanishes for
(i)~a depolarizing channel with $p=1/4$ ($p_x=p_y=p_z=1/12$)~\cite{bib_bruss}; 
(ii)~a ``2-Pauli'' channel with $p=1/3$ ($p_x=p_z=1/6$, $p_y=0$);
and (iii)~a dephasing channel with $p=1/2$ ($p_x=p_y=0$, $p_z=1/2$).
Furthermore, using the fact that $C$ cannot be
superadditive for a convex combination of a perfect 
and a noisy channel~\cite{bib_bdsw}, an upper bound on $C$
can be written using a linear interpolation between the perfect channel
$(0,0,0)$ and any Pauli channel lying on $E$:\footnote{It is worth
noting that the proof in~\cite{bib_bdsw} is not valid at the limit
of a noisy channel of vanishing capacity, which happens to be the case
on $E$. Therefore, the resulting upper bound on $C$ is
rigorously proven only if the additional assumption is made that $C$
is a continuous function of $p_x$, $p_y$ and $p_z$.}
\begin{equation}  \label{eq_upperbound}
C \le 1-2(x^2+y^2+z^2+xy+xz+yz) \; .
\end{equation}
Note that another class of symmetric PCMs can be found by
requiring $\rho_{RA}=\rho_{RC}$, i.e., considering $C$ as the second
output and $B$ as the cloning machine. This requirement 
implies $v=x-y+z$
rather than Eq.~(\ref{eq_condition-dupl}), which gives rise
to the reflection of $E$ with respect to the $xz$-plane, i.e, $y\to -y$. 
It does not change the above bound on $C$ because this class of PCMs
has noisier outputs in the first octant $x,y,z\ge 0$.
\par

\subsection{Quantum triplicators based on the PCM}

Let us turn to the fully symmetric PCMs that have {\em three} outputs
emerging from the {\em same} Pauli channel, i.e.,
$\rho_{RA}=\rho_{RB}=\rho_{RC}$, which corresponds
to a family of (non-optimal) quantum {\em triplicating} machines.
The requirement $\rho_{RA}=\rho_{RC}$ implies $v=x-y+z$, 
which, together with Eq.~(\ref{eq_condition-dupl}),
yields the conditions
\begin{equation}
(v=x+z) \wedge (y=0) \;.
\end{equation}
Incidentally, we notice
that if {\em all} pairs are required to be in the {\em same}
mixture of Bell states, this mixture cannot have
a singlet $|\Psi^-\rangle$ component. The outputs of the corresponding 
triplicators emerge therefore from a Pauli channel with $p_y=0$,
so that these triplicators are {\em state-dependent}, in contrast with
the one considered in Ref.~\cite{bib_gisinmassar}.\footnote{For describing
a {\em state-independent} triplicator, a 6-qubit wave function should be
used, that is, the cloner should consist of 2 qubits.}
These triplicators are represented by the
intersection of $E$ with the $xz$-plane, that is, the ellipse
\begin{equation}
x^2+z^2+xz={1\over 2}  \; ,
\end{equation}
whose semiminor axis is $1/\sqrt{3}$ [oriented along the direction
$(1,1)$] and semimajor axis is $1$. The intersection of this ellipse
with its semiminor axis ($x=z=1/\sqrt{6}$) corresponds to
the 4-qubit wave function
\begin{equation}  \label{eq_psi-triplicator}
|\Psi\rangle_{RABC}={2\over\sqrt{6}}|\Phi^+\rangle|\Phi^+\rangle
+{1\over\sqrt{6}}|\Phi^-\rangle|\Phi^-\rangle
+{1\over\sqrt{6}}|\Psi^+\rangle|\Psi^+\rangle  \; ,
\end{equation}
which is symmetric under the interchange of any two qubits
and maximizes the 2-bit entropy (or minimizes the mutual entropy between
any two outputs of the triplicator, making them maximally
independent). Equation~(\ref{eq_psi-triplicator}) thus
characterizes the best triplicator of this ensemble,
whose three outputs emerge from
a ``2-Pauli'' channel with $p=1/3$ ($p_x=p_z=1/6$). 
According to Eq.~(\ref{eq_2Paulichannel}), the (state-dependent)
operation of this triplicator on an arbitrary qubit can be written as
\begin{equation} \label{eq_best-triplicator}
|\psi\rangle \to {1\over 2}|\psi\rangle\langle\psi|
+ {1\over 6}|\psi^*\rangle\langle\psi^*|
+ {1 \over 3} (\openone /2)  \; .
\end{equation}
which reduces to the triplicator 
that was considered in Ref.~\cite{bib_bh_tripl}. 
Note that if $|\psi\rangle$ is real, Eq.~(\ref{eq_best-triplicator})
reduces to Eq.~(\ref{eq_UCMstandout}), so that 
the three outputs are the same as those of the UCM.
The fidelity of cloning is then the same as for the UCM ($f=5/6$)
regardless the input state (provided it is real).
\par

\section{Heisenberg cloning machines for $N$-dimensional states}
\label{sect_Ndim}

\subsection{Channel characterization using the maximally-entangled states}

Consider now the cloning of the state of an $N$-dimensional
system. In order to follow our previous discussion for quantum bits ($N=2$),
we need first to generalize the Bell states
and introduce a set of $N^2$ maximally-entangled (ME)
states of two $N$-dimensional systems, $A$ and $B$:
\begin{equation}
|\psi_{m,n}\rangle_{AB} = {1\over\sqrt{N}} \sum_{j=0}^{N-1}
{\rm e}^{2\pi i (j n / N)} |j\rangle_A |j+m \rangle_B 
\end{equation}
where the indices $m$ and $n$ ($m,n=0,\cdots, N-1$) 
label the $N^2$ states. Note that, here and below, the ket labels
are taken modulo $N$. Taking the partial trace of any state
$|\psi_{m,n}\rangle \langle \psi_{m,n}|$ results
in a density operator for $A$ or $B$ given by
\begin{equation}
\rho_A=\rho_B={1\over N} \sum_{j=0}^{N-1} |j\rangle\langle j| =
\openone /N
\end{equation}
implying that $A$ and $B$ are maximally entangled.
It is easy to check that the $|\psi_{m,n}\rangle$ are orthonormal
and form a complete basis in the product Hilbert spaces
${\cal H}_A\otimes {\cal H}_B$. The resolution of identity simply reads
\begin{eqnarray}
\sum_{m,n=0}^{N-1} 
|\psi_{m,n}\rangle \langle\psi_{m,n}|
 &=& {1\over N} \sum_{m,n}\sum_{j,j'} {\rm e}^{2\pi i [(j-j') n / N]} 
     |j\rangle\langle j'|  \otimes |j+m\rangle \langle j'+m|
     \nonumber\\
 &=& {1\over N} \sum_{k,n}\sum_{j,j'} {\rm e}^{2\pi i [(j-j') n / N]} 
     |j\rangle\langle j'|  \otimes |k\rangle \langle k+j'-j|
     \nonumber\\
 &=& \sum_{k,j}
     |j\rangle\langle j|  \otimes |k\rangle \langle k|
     \nonumber\\
 &=& \openone_A \otimes \openone_B
\end{eqnarray}
where we have made the substitution $j+m=k$ and used the identity
\begin{equation}
\sum_{n=0}^{N-1} {\rm e}^{2\pi i [(j-j') n / N]} = N \, \delta_{j,j'}
\end{equation}
The maximally-entangled (ME) states $|\psi_{m,n}\rangle$
generalize the Bell states for $N>2$:
in the special case of two maximally-entangled qubits ($N=2$),
we simply have the equivalence $|\psi_{0,0}\rangle=|\Phi^+\rangle$,
$|\psi_{0,1}\rangle=|\Phi^-\rangle$, $|\psi_{1,0}\rangle=|\Psi^+\rangle$, and
$|\psi_{1,1}\rangle=|\Psi^-\rangle$.
\par

We now describe a quantum Heisenberg channel that processes
$N$-dimensional states by using the correspondence between
these ME-states and the (Heisenberg) group of error operators $U_{m,n}$
on a $N$-dimensional state~\cite{bib_fivel}. In such a channel,
an arbitrary state $|\psi\rangle$ undergoes a particular
unitary transformation (or error)
\begin{equation}
U_{m,n} = \sum_{k=0}^{N-1} {\rm e}^{2\pi i (k n / N)}
|k+m\rangle\langle k|
\end{equation}
with probability $p_{m,n}$ (with $\sum_{m,n} p_{m,n}=1$).
Note that $U_{0,0}=\openone$, implying that $|\psi\rangle$ is left
unchanged with probability $p_{0,0}$.
These error operators generalize the Pauli matrices for qubits:
$m$ labels the ``shift'' errors (generalizing the bit flip $\sigma_x$)
while $n$ labels the phase errors (generalizing the phase flip $\sigma_z$).
If the input of the channel $X$ is maximally entangled
with a reference $R$ (an $N$-dimensional system)
so that their joint state is
$|\psi_{0,0}\rangle=\sum_j |j\rangle |j\rangle / \sqrt{N}$, then
the joint state of the output $Y$ and $R$ is simply
a mixture of the $N^2$ ME-states,
\begin{equation}
\rho_{RY}=\sum_{m,n} p_{m,n} |\psi_{m,n}\rangle\langle\psi_{m,n}|
\end{equation}
generalizing the mixture of Bell states that we had for qubits
in Sec.~\ref{sect_paulichannel}.
Indeed, applying $U_{m,n}$ {\em locally} (i.e., to one subsystem, 
leaving the other unchanged) transforms $|\psi_{0,0}\rangle$ into a ME-state,
\begin{equation}
(\openone\otimes U_{m,n})|\psi_{0,0}\rangle = |\psi_{m,n}\rangle
\end{equation}
extending Eq.~(\ref{eq_actionPauli}) to $N>2$.
This allows us to treat the cloning of $N$-dimensional states 
following closely Sec.~\ref{sect_PCM}, that is, by
considering a 4-partite pure state, namely the state
of a reference system $R$ (initially 
entangled with the input $X$), the two outputs $A$ and $B$, and
the cloning machine (or a third output) $C$.
Note that, using the same reasoning as in Sec.~\ref{sect_PCM}, 
it appears that the minimum size required for the Hilbert space 
of the cloning machine is $N$. In order to purify $\rho_{RA}$,
we need a $N^2$-dimensional additional space whereas $B$ is only
$N$-dimensional, so that an additional $N$-dimensional cloner
is necessary. As for the Pauli cloning machine, we conjecture
that it is sufficient. Consequently, 
we need to consider a pure state in a $N^4$-dimensional
Hilbert space in order to characterize the entire set
of $N$-dimensional Heisenberg cloning machines.
\par

\subsection{Asymmetric Heisenberg cloning machines}

We start by expressing the joint state of the four
$N$-dimensional systems $R$, $A$, $B$, and $C$, as
a superposition of double-ME states:
\begin{equation}   \label{eq_psiRABX}
|\Psi\rangle_{RA;BC} =
\sum_{m,n=0}^{N-1} \alpha_{m,n} \; |\psi_{m,n}\rangle_{RA} \;
|\psi_{m,N-n}\rangle_{BC}
\end{equation}
where the $\alpha_{m,n}$ are (arbitrary) complex amplitudes
such that $\sum_{m,n}| \alpha_{m,n}|^2 =1$.
This expression reduces to Eq.~(\ref{eq_psi}) for $N=2$.
By tracing the state Eq.~(\ref{eq_psiRABX}) over $B$ and $C$,
we see that the joint state of $R$ and $A$
is a mixture of the ME-states, 
\begin{equation}
\rho_{RA}= \sum_{m,n=0}^{N-1} |\alpha_{m,n}|^2 \;
           |\psi_{m,n}\rangle \langle\psi_{m,n}|
\end{equation}
so that $A$ can be viewed as the
output of a Heisenberg channel that processes an input maximally entangled
with $R$ (the initial joint state being $|\psi_{0,0}\rangle$).
Now, we will show that, by interchanging $A$ and $B$,
the joint state of the 4-partite system can be reexpressed as
a superposition of double-ME states
\begin{equation}   \label{eq_psiRBAX}
|\Psi\rangle_{RB;AC} =
\sum_{m,n=0}^{N-1} \beta_{m,n} \; |\psi_{m,n}\rangle_{RB} \;
|\psi_{m,N-n}\rangle_{AC}
\end{equation}
where the amplitudes $\beta_{m,n}$ are defined by
\begin{equation}  \label{eq_defbeta}
\beta_{m,n} = {1\over N} \sum_{x,y=0}^{N-1} 
{\rm e}^{2\pi i [(nx-my) / N]} \; \alpha_{x,y}
\end{equation}
These amplitudes characterize the quantum channel leading to
the second output, $B$, since the joint state of $R$ and $B$
is again a mixture of ME-states,
\begin{equation}
\rho_{RB}= \sum_{m,n=0}^{N-1} |\beta_{m,n}|^2 \;
           |\psi_{m,n}\rangle \langle\psi_{m,n}|
\end{equation}
Thus, outputs $A$ and $B$ of the $N$-dimensional cloning machine
emerge from channels of respective probabilities $p_{m,n}=|\alpha_{m,n}|^2$
and $q_{m,n}=|\beta_{m,n}|^2$ which are related via
Eq.~(\ref{eq_defbeta}). It is easy to check that Eq.~(\ref{eq_defbeta})
reduces to Eq.~(\ref{eq_DFT-qubits}) for qubits ($N=2$).
\par

Let us prove Eqs.~(\ref{eq_psiRBAX}) and (\ref{eq_defbeta})
by considering a single component
$|\psi_{\mu,\nu}\rangle_{RA} |\psi_{\mu,N-\nu}\rangle_{BX}$
in Eq.~(\ref{eq_psiRABX}), that is,
choosing $\alpha_{m,n}= \delta_{m,\mu} \delta_{n,\nu}$.
Then, Eq.~(\ref{eq_defbeta}) 
gives $\beta_{m,n}= {\rm e}^{2\pi i [(n\mu-m\nu) / N]} /N$,
so that Eq.~(\ref{eq_psiRBAX}) results in
\begin{eqnarray}
 |\Psi\rangle_{RB;AC} &=& {1 \over N}
\sum_{m,n}  {\rm e}^{2\pi i [(n\mu-m\nu) / N]} \;
|\psi_{m,n}\rangle_{RB} \; |\psi_{m,N-n}\rangle_{AC}
\nonumber \\
&=& {1\over N^2} \sum_{m,n} \sum_{j,j'}
{\rm e}^{2\pi i [(n\mu-m\nu) / N]} \;
{\rm e}^{2\pi i [(j-j') n / N]} \;
     |j\rangle_R \, |j+m\rangle_B \, |j'\rangle_A \, |j'+m\rangle_C
\nonumber \\
&=& {1\over N} \sum_{m,j} {\rm e}^{-2\pi i (m\nu / N)}
     |j\rangle_R \, |j+m\rangle_B \, |j+\mu\rangle_A \, |j+\mu+m\rangle_C
\end{eqnarray}
where we have used $\sum_n {\rm e}^{2\pi i [(\mu+j-j')n / N]}
=N\,\delta_{j+\mu,j'}$. Making the substitution $k=j+m$, we obtain
\begin{equation}
 |\Psi\rangle_{RB;AC} = {1 \over N}
\sum_{j,k} {\rm e}^{2\pi i [(j-k)\nu/ N]}
     |j\rangle_R \, |k\rangle_B \, |j+\mu\rangle_A \, |k+\mu\rangle_C
\end{equation}
which is indeed equivalent to 
$|\psi_{\mu,\nu}\rangle_{RA} |\psi_{\mu,N-\nu}\rangle_{BX}$
when interchanging $A$ and $B$. This proof holds for an arbitrary
$\alpha_{m,n}$ as a consequence of the 
linearity of Eq.~(\ref{eq_defbeta}).
\par

The latter equation is basically a 2-dimensional discrete Fourier
transform (up to an interchange of the indices $m$ and $n$, and a minus sign):
\begin{equation}
\beta_{m,n}=F[n,m] \qquad{\rm with~}
F[{\tilde x},{\tilde y}]= {\cal F}_2 \{ \alpha_{N-x,y} \}
\end{equation}
where ${\cal F}_2$ is a 2-dimensional discrete Fourier transform.
The normalization of the $\beta_{m,n}$'s simply results from
Parseval's theorem: $\sum_{m,n} |\alpha_{m,n}|^2=\sum_{m,n}
|\beta_{m,n}|^2$. Therefore, we have shown that
the {\em complementarity} between the two outputs $A$ and $B$
of an $N$-dimensional Heisenberg cloning machine 
is simply governed by the relationship
between a function and its Fourier transform. 
This emphasizes that, if one output is close-to perfect
($\alpha_{m,n}$ is a peaked function), then the second one is
very noisy ($\beta_{m,n}$ is a flat function), and conversely.
In other words, the indices of $\alpha_{m,n}$ and $\beta_{m,n}$
act as conjugate variables, so that the probability distributions
characterizing the two outputs, $p_{m,n}$ and $q_{m,n}$, 
cannot have a variance simultaneously tending to zero.
(Note that the index $m$ of $p_{m,n}$ is dual to the index $n$ of
$q_{m,n}$, and conversely.)
A symmetric $N$-dimensional HCM
then corresponds simply to a function $\alpha_{m,n}$ whose square 
is equal to its squared Fourier transform, i. e.,
$|\alpha_{m,n}|^2=|\beta_{m,n}|^2$. 
\par

\subsection{No-cloning inequality for $N$-dimensional states}
\label{sect_Ndim_nocloning}

We now investigate this complementarity principle
in the special case of {\em isotropic} HCMs. Thus,
the channel underlying each output
is a {\em depolarizing} channel, that is, all the probabilities
$p_{m,n}$ are equal except $p_{0,0}$ (and equivalently for $q_{m,n}$).
Assume that $\alpha_{m,n}$ is the superposition of a peaked component
$P_{m,n} = \delta_{m,0} \, \delta_{n,0}$ (i.~e., a perfect channel)
and a flat component $F_{m,n} = 1/N$ (i.~e., a fully depolarizing channel),
with respective amplitudes $\hat{a}$ and $a$:
\begin{equation}
\alpha_{m,n}= \hat{a} \, P_{m,n} + a \, F_{m,n}
\end{equation}
Note that the normalization condition
$|\overline{a}+a/N|^2+(N^2-1)|a/N|^2 =1$ can be written as
\begin{equation}
|\overline{a}|^2 + {2\over N} \, {\rm Re}(\overline{a}a^*) + |a|^2=1
\end{equation}
Tracing over $B$ and $C$, we see that the first output 
is characterized by
\begin{equation}
\rho_{RA}= \left[ |{\overline a}|^2 + {2\over N} \, {\rm Re}({\overline a} a^*)
\right]  |\psi_{0,0}\rangle\langle\psi_{0,0}|
+ |a|^2 \; {\openone\otimes \openone \over N^2}
\end{equation}
so that the input state is replaced by a random state with probability
$\pi_a=|a|^2$ and left unchanged with probability $1-\pi_a$.
This is the $N$-dimensional generalization of a depolarizing channel:
if $a=0$, the channel is perfect, while $a=1$ corresponds to a fully
depolarizing channel. Thus, $\pi_a$ denotes the {\em depolarization}
fraction of the channel, while $s_a=1-\pi_a$
is the scaling factor for output $A$.
Using Eq.~(\ref{eq_defbeta}), we see that
the second output is characterized by
\begin{equation}
\beta_{m,n} = \hat{b} \, P_{m,n}+ b \, F_{m,n}
\end{equation}
where $\hat{b}=a$ and $b=\hat{a}$
since $F_{m,n}$ are $P_{m,n}$ are dual under Fourier transform.
Here $\pi_b=|b|^2$ is the depolarizing fraction of the 
channel associated with $B$.
Thus, the complementarity of the two outputs of the class
of (asymmetric) isotropic cloners considered here 
can be simply written as
\begin{equation}  \label{eq_N_nocloneqal}
|a|^2 + {2\over N} {\rm Re}(ab^*) +|b|^2 =1
\end{equation}
It is easy to see that the best cloning (the smallest values
for $|a|$ and $|b|$) is achieved when the cross term 
is the largest in magnitude, that is, when
$a$ and $b$ have the same phase. For simplicity, we assume
that $a$ and $b$ are real and positive.
Therefore, arguing like before, we find a {\em no-cloning} inequality
for an $N$-dimensional quantum state:
\begin{equation} \label{eq_N_noclonineq}
a^2 + {2 \over N}\, ab +b^2 \ge 1
\end{equation}
where $\pi_a=a^2$ or $\pi_b=b^2$ are the depolarizing fractions underlying
outputs $A$ and $B$, respectively.
This corresponds to the domain in the $(a,b)$-space which is outside
an ellipse, oriented just as in Fig.~\ref{fig_nocloning}, 
whose semiminor axis is $\sqrt{N/(N+1)}$
and semimajor axis is $\sqrt{N/(N-1)}$.
Equation~(\ref{eq_N_noclonineq}) generalizes the no-cloning
inequality for qubits, Eq.~(\ref{eq_no_clon_uncert}),
which is simply equivalent to Eq.~(\ref{eq_N_noclonineq}) for $N=2$.
This ellipse intercepts its minor axis at $(\sqrt{N/2(N+1)},\sqrt{N/2(N+1)})$,
which corresponds to an $N$-dimensional UCM~\cite{bib_bh_registers},
as discussed below.
\par

Note that this ellipse tends to a circle of radius one
as $N$ tends to infinity. This means that,
at the limit $N\to \infty$,  the sum of the depolarizing fractions
cannot be lower than one, i.e., $\pi_a+\pi_b\ge 1$.
The no-cloning inequality involves then an ``incoherent'' sum in this limit
(i.e., probabilities---not amplitudes---are added, while the
cross term disappears), which emphasizes that $N\to\infty$ can be viewed
as a semi-classical limit. The optimal
cloning machine (with $\pi_a+\pi_b=1$) can then be understood in
classical terms: the input state is sent to output $A$ or $B$
with probability $1-\pi_a=\pi_b$ or $1-\pi_b=\pi_a$, respectively, 
the other output being a random
$N$-dimensional state. There is no such classical interpretation 
for finite-$N$ cloners, as $(1-\pi_a)+(1-\pi_b)$ can then exceed one.
For example, for qubits ($N=2$), we have $\pi_a=\pi_b=1/3$, so that
the input qubit is apparently sent to each output with probability
$2/3$, which makes a total of 4/3 (!). 
\par

\subsection{Entropic no-cloning uncertainty relation}
\label{sect_entropicnocloning}

As mentioned earlier, the tradeoff between the quality
of the two copies is the consequence of an ``uncertainty principle'' 
inherent to Fourier transforms. We can express this uncertainty principle
in general 
by making use of the entropic uncertainty relations for non-commuting
observables~\cite{bib_deutsch,bib_kraus,bib_maassen}. Consider two
observables $O_A$ and $O_B$ whose respective set of eigenvectors
are $\{|a_j\rangle\}$ and $\{|b_k\rangle\}$. For any quantum
state $|\psi\rangle$, the probability distributions
\begin{eqnarray}
p_j&=& |\langle a_j|\psi\rangle| \nonumber \\
q_k&=& |\langle b_k|\psi\rangle|
\end{eqnarray}
associated with the measurement of $O_A$ and $O_B$
cannot be peaked simultaneously if $O_A$ and $O_B$ do not commute.
The uncertainty associated with $p_j$ and $q_k$ can be measured
by using the Shannon entropies $H[p_j]=-\sum_j p_j \log_2 p_j$ and
$H[q_k]=-\sum_k q_k \log_2 q_k$. 
Then, the entropic inequality
\begin{equation}  \label{eq_deutsch}
H[p_j] + H[q_k] \ge -2 \log_2 (c)
\qquad {\rm with~}c=\max_{j,k}|\langle a_j|b_k\rangle|
\end{equation}
can be shown to hold for any state $|\psi\rangle$, thereby
expressing the balance between the uncertainty
of the measurement of $O_A$ and $O_B$.
Equation~(\ref{eq_deutsch}) can be applied to cloning by
considering observables $O_A$ and $O_B$
with respective eigenvectors $|m,n \rangle$
and $|\psi_{m,n}\rangle$, resulting in the
entropic no-cloning uncertainty relation
\begin{equation}  \label{eq_entropic_nocloning}
H[p_{m,n}] + H[q_{m,n}] \ge \log_2(N^2)
\end{equation}
Thus, the sum of the entropies of the probability distribution of the $N^2$
error operators affecting each of the two copies cannot be less
than $\log_2(N^2)$. The bound is saturated when one copy is perfect
(vanishing entropy) as the other copy corresponds then to a flat 
distribution (maximum entropy).
Note that Eq.~(\ref{eq_entropic_nocloning}) is a special case
of the entropic no-cloning inequality
derived in Ref.~\cite{bib_cerf_nasa} which applies to {\em any} 
cloning machine: $L_A+L_B\ge 2S$ where $L_A$ and $L_B$ are the
{\em losses}~\cite{bib_adami} of the channels leading
to the two outputs of the cloner while $S$ is the 
source entropy.\footnote{In this paper, we restrict ourselves 
to cloners whose outputs emerge from Heisenberg channels. In that
case, it is easy to show
that $H[p_{m,n}]$ and $H[q_{m,n}]$ are simply the 
losses $L_A$ and $L_B$ of the channels leading
to the outputs of the HCM.} 
Equation~(\ref{eq_entropic_nocloning}) is unfortunately not a tight bound
as can be seen for the UCM for qubits ($N=2$): we have indeed
$H[p_{m,n}]=H[q_{m,n}]=2-\log_2(3)/2=1.21$~bits, so that
\begin{equation}
H[p_{m,n}] + H[q_{m,n}]=2.42>2
\end{equation}
\par

Alternate entropic no-cloning uncertainty relations can also be obtained
by exploiting the fact that $p_{m,n}$ and $q_{m,n}$ are
related by a 2-dimensional Fourier
transform. Since the index $m$ of $p_{m,n}$ is dual to the index
$n$ of $q_{m,n}$, we can use Eq.~(\ref{eq_deutsch})
with eigenvectors $|m,0\rangle$ and $N^{-1/2} \sum_m |\psi_{m,n}\rangle$,
which results in
\begin{equation}  \label{eq_entruncer1}
H[p_m] + H[q_n] \ge \log_2(N)
\end{equation}
with $p_m=\sum_n p_{m,n}$ and $q_n=\sum_m q_{m,n}$.
Conversely, we have
\begin{equation}  \label{eq_entruncer2}
H[p_n] + H[q_m] \ge \log_2(N)
\end{equation}
with $p_n=\sum_m p_{m,n}$ and $q_m=\sum_n q_{m,n}$.
For the UCM for qubits, $p_m$, $p_n$, $q_m$, and $q_n$
are all equal to (5/6,1/6), so that
$H[p_n]=H[p_m]=H[q_n]=H[q_m]=0.65$~bits, implying that
Eqs.~(\ref{eq_entruncer1}) and (\ref{eq_entruncer2})
are not saturated.
\par

\subsection{Symmetric cloning machine or the $N$-dimensional UCM}

It is easy to find the {\em symmetric} $N$-dimensional
cloner of the class of isotropic HCMs (i.e., cloners
whose outputs emerge from an $N$-dimensional depolarizing channel)
by requiring that $a=b$ in Eq.~(\ref{eq_N_nocloneqal}),
which simply results in a depolarizing fraction
\begin{equation}
\pi=|a|^2 = { N \over 2(N+1)}
\end{equation}
The underlying 4-partite wave function for the reference, 
the two outputs, and the cloner is
\begin{equation}  \label{eq_4-partite-wf}
|\Psi\rangle_{RA;BC} = 
\hat{a} \; |\psi_{0,0}\rangle_{RA} |\psi_{0,0}\rangle_{BC}
+ {a\over N} \sum_{m,n=0}^{N-1}  
|\psi_{m,n}\rangle_{RA} |\psi_{m,N-n}\rangle_{BC}
\end{equation}
with $a=\hat{a}=\sqrt{N\over 2(N+1)}$.
Using $\sum_{m,n} |\psi_{m,n}\rangle |\psi_{m,N-n}\rangle
=\sum_{j,k}|j\rangle |k\rangle |j\rangle |k\rangle$,
we find that Eq.~(\ref{eq_4-partite-wf}) can be rewritten as
\begin{equation}
|\Psi\rangle_{RA;BC} = 
\sqrt{1\over 2N(N+1)} \sum_{j,k=0}^{N-1}
\big( |j\rangle_R |j\rangle_A |k\rangle_B |k\rangle_C +
|j\rangle_R |k\rangle_A |j\rangle_B |k\rangle_C \big)
\end{equation}
which immediately implies that it is symmetric under the
interchange of $A$ and $B$, as expected.
By tracing over $B$ and $C$, Eq.~(\ref{eq_4-partite-wf}) yields
\begin{equation}
\rho_{RA}=\rho_{RB}=
{N+2\over 2(N+1)}\; |\psi_{0,0}\rangle\langle\psi_{0,0}|
+ {1\over 2N(N+1)} \sum_{m,n=0}^{N-1}  
|\psi_{m,n}\rangle\langle\psi_{m,n}|
\end{equation}
which shows that this cloner is
state-independent since it acts on an arbitrary state as
\begin{equation}
|\psi\rangle \to \rho={N+2 \over 2(N+1)} \; |\psi\rangle\langle\psi|
+ { N \over 2(N+1)} \; (\openone/N)
\end{equation}
Thus, the scaling factor corresponding to both outputs
is given by
\begin{equation}
s=1-|a|^2= {N+2 \over 2(N+1)}
\end{equation}
in agreement with the expression
derived in Refs.~\cite{bib_bh_registers,bib_werner}
for the $N$-dimensional UCM. 
When $N\to \infty$, the UCM can be viewed as a
classical machine that is transmitting the input state to one
of the two outputs with probability 1/2, a random state being sent
on the other output.
\par

In analogy with what we have done for quantum bits ($N=2$) 
in Sect.~\ref{sect_symmPCM}, it should
be possible to find an entire class of symmetric cloners with $N>2$,
thereby generalizing Eq.~(\ref{eq_ellipsoid}). Roughly speaking,
this class should be based on functions that are equal (in magnitude)
to their Fourier transform. This should give rise
to an upper bound on the quantum capacity of 
a Heisenberg channel processing $N$-dimensional states, 
extending the bound Eq.~(\ref{eq_upperbound}) for Pauli channels. 
This will be investigated elsewhere.

\section{Conclusion}

We have defined a class of {\em asymmetric} cloners
for quantum bits (Pauli cloning machines) 
and $N$-dimensional quantum states (Heisenberg cloning machines).
For quantum bits, we have shown that the PCM, 
whose outputs emerge from two non-identical Pauli channels, 
generalizes the universal cloning machine 
of Buzek and Hillery~\cite{bib_bh}. 
The class of isotropic (but asymmetric) PCMs allowed us
to derive a tight no-cloning inequality for quantum bits,
quantifying the impossibility of copying due to quantum mechanics.
Using a class of symmetric (but anisotropic) PCMs,
we also established an upper bound on the quantum
capacity of the Pauli channel.
These considerations have been extended to $N$ dimensions,
showing that the notion of asymmetric cloners is quite general.
We have defined the $N$-dimensional HCM, whose outputs
emerge from two distinct Heisenberg channels.
The $N$-dimensional universal cloning 
machine~\cite{bib_bh_registers,bib_werner} appears as a special
case---symmetric and isotropic---of these cloners.
Using isotropic (asymmetric) HCMs,
we have generalized the no-cloning inequality in order to characterize
the impossibility of perfectly copying $N$-dimensional states.
Furthermore, we have shown that the tradeoff governing 
the quality of the two copies of an $N$-dimensional state
results from an {\em uncertainty principle} 
akin to the complementarity between position and momentum,
implying that 
the probability distributions of the error operators
affecting each copy are just the square of two dual functions 
under a Fourier transform.

\acknowledgements

I am grateful to V. Buzek and A. Peres for very helpful
discussions. This work was supported in part by the NSF
under Grant Nos. PHY 94-12818 and PHY 94-20470,
and by a grant from DARPA/ARO through the QUIC Program
(\#DAAH04-96-1-3086).

\end{document}